\documentclass[prl,twocolumn,showpacs,preprintnumbers,amsmath,amssymb]{revtex4}

\usepackage{graphicx}
\usepackage{dcolumn}
\usepackage{bm}

\begin{document}

\newcommand*{\cm}{cm$^{-1}$\,}
\newcommand*{\Tc}{T$_c$\,}


\title{Origin of the structural phase transition at 130 K in BaNi$_2$As$_2$: a combined study of
optical spectroscopy and band structure calculations}
\author{Z. G. Chen}
\author{G. Xu}
\author{W. Z. Hu}
\author{X. D. Zhang}
\author{P. Zheng}
\author{G. F. Chen}
\author{J. L. Luo}
\author{Z. Fang}
\author{N. L. Wang}

\affiliation{Beijing National Laboratory for Condensed Matter
Physics, Institute of Physics, Chinese Academy of Sciences,
Beijing 100190, China}


\begin{abstract}

BaNi$_2$As$_2$ exhibits a first order structural transition at 130
K. Understanding this structural transition is a crucial step
towards understanding the electronic properties of the material.
We present a combined optical spectroscopy and band structure
calculation study on the compound across the transition. The study
reveals that BaNi$_2$As$_2$ is a good metal with a rather high
plasma frequency. The phase transition leads to a small reduction
of conducting carriers. We identify that this reduction is caused
by the removal of several small Fermi surface sheets contributed
dominantly from the As-As bonding and Ni-As antibonding.

\end{abstract}



\maketitle

The discoveries of superconductivity in Fe- or Ni-based layered
pnictide compounds\cite{Kamihara06,Watanabe,Kamihara08} have
generated tremendous interest in condensed matter community. For
the iron arsenic-based compounds, not only could they have high
superconducting transition temperatures\cite{ChenXH,Chen1,Ren1}
but also show intriguing interplay between structure, magnetism
and superconductivity. The undoped compounds commonly display the
structural and magnetic phase transitions, which, depending on
materials, could occur either at the same temperature or
separately\cite{Cruz,Rotter2,Chu,Chen2}. Upon electron or hole
doping or application of pressure, both the magnetic order and the
structural transition are suppressed, and superconductivity
emerges\cite{JZhao,Rotter2,Torikachvili}. Intensive studies have
been done to elucidate the origin of the structural/magnetic
transitions and their relation to superconductivity. The magnetic
transitions were identified as itinerant spin-density-wave (SDW)
instabilities caused largely by the strong nesting tendency
between the hole and electron Fermi surfaces\cite{Dong,Singh},
though conflicting views from a local superexchange picture also
exist. The structural phase transition was widely believed to be
driven by the magnetic transition\cite{Yildirim,FangXu}.

Compared with the FeAs-based compounds, the NiAs-based systems
were much less studied. Superconductivity in Ni-pnictides has been
found in quaternary ZrCuSiAs structure type LaNiPO (T$_c$=4
K)\cite{Watanabe}, LaNiAsO (T$_c$=2.7 K)\cite{Li} and
ThCr$_2$Si$_2$ structure type ANi$_2$As$_2$ (A=Ba, Sr)
(T$_c$=0.6$\sim$0.7 K)\cite{Ronning,Bauer,Sefat}. Although the
Ni-pnictide superconductors share the same Ni$_2$As$_2$ PbO-type
structure as Fe-pnictides, the superconducting transition
temperatures are much lower, never exceeds 5 K even upon
doping\cite{Li}. Nevertheless, a first order phase transition at
T$_s$=130 K, much similar to the AFe$_2$As$_2$ (A=Ca, Sr, Ba), is
found for BaNi$_2$As$_2$ compound\cite{Ronning,Sefat}. Initially,
by analogy with FeAs-based 122 compounds, the transition was
considered as a magnetic SDW transition concomitant with the a
structural transition\cite{Ronning}. However, to date no evidence
for magnetic transition was reported. Much remain unknown about
this transition. It is highly interesting to study the origin of
the first order phase transition and compare with their FeAs
analogies.

In this work we report the first optical spectroscopy study on
BaNi$_2$As$_2$ single crystals. The study reveals remarkable
differences between BaNi$_2$As$_2$ and BaFe$_2$As$_2$. In accord
with the presence of two more electrons in 3d orbitals, the
BaNi$_2$As$_2$ shows a much higher plasma frequency (or carrier
density). By lowering the temperature across the transition, part
of the spectral weight from conducting electrons was removed and
shifted to high energies. However, unlike the case of
BaFe$_2$As$_2$, where an SDW gap opens at low
frequencies\cite{Hu}, the spectral weight redistribution for
BaNi$_2$As$_2$ occurs over a much broader energy scale. The study
reveals that the first order phase transition in BaNi$_2$As$_2$ is
essentially different from the SDW type phase transition as
observed in BaFe$_2$As$_2$, thus against the presence of competing
phase such as magnetic SDW order in Ni-pnictides. To understand
the spectral change across the transition, we performed
first-principle calculations of the band structure. We illustrate
that the transition is mainly caused by the removal of several
small Fermi surface sheets with dominant As-As $pp\sigma$ bonding
and Ni-As $dp\sigma*$ anitbonding character. The other
orbitals/bands are less affected and remains conductive.

Single crystals of BaNi$_2$As$_2$ were grown using Pb-flux method
similar to the procedure described by Ronning et
al.\cite{Ronning}, except that we use a much slowly cooling rate
of 3.5 $^o$C from 1100 $^o$C to 650 $^o$C. The Pb flux were
removed by using a centrifuge at 650 $^o$C. Resistivity
measurement, as shown in the inset of Fig. 1, indicates metallic
behavior with a small resistivity value of about 50 $\mu\Omega cm$
at 300 K. The resistivity shows a jump at 130 K on cooling, then
becomes more metallic at low temperature, similar to the case of
CaFe$_2$As$_2$. Finally, BaNi$_2$As$_2$ enters into
superconducting state at 0.65 K. All those are in good agreement
with previous reports in literature\cite{Ronning,Sefat}.

\begin{figure}[t]
\includegraphics[width=8.5cm]{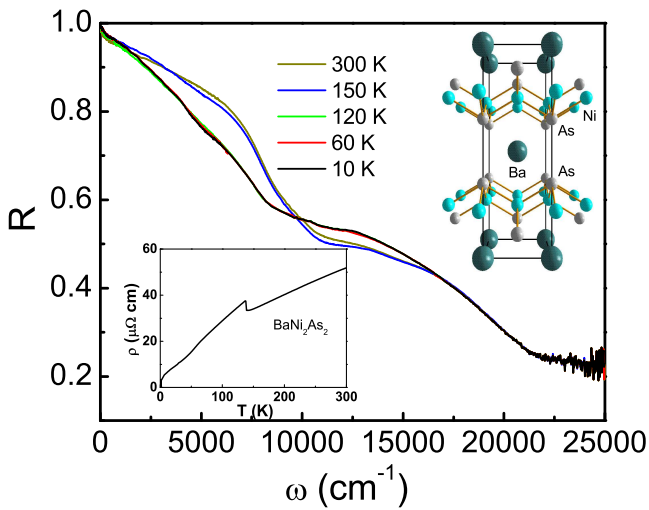}
\caption{\label{fig:R}(Color online) Reflectance curves up to
25000 \cm ($\sim$3 eV) at different temperatures. Upper inset
shows the crystal structure. Lower inset shows the T-dependent dc
resistivity.}
\end{figure}

The optical reflectance measurements were performed on a
combination of Bruker IFS 66v/s and 113v on newly cleaved surfaces
(ab-plane) of those crystals up to 25000 cm$^{-1}$. An \textit{in
situ} gold and aluminium overcoating technique was used to get the
reflectivity R($\omega$). The real part of conductivity
$\sigma_1(\omega)$ is obtained by the Kramers-Kronig
transformation of R($\omega$).

Figure 1 shows the experimental reflectance spectra up to 25000
\cm. A rather good metallic response is observed. R($\omega$)
approaches to unity at zero frequency, and shows a plasma
edge-like shape near 10000 \cm. Note that this edge frequency is
substantially higher than that in BaFe$_2$As$_2$, which is seen
around only 3000 \cm where it merges into the relatively high
values of mid-infrared reflectance contributed mainly by the
interband transitions\cite{Hu}. This gives direct evidence that
the conducting carrier density in BaNi$_2$As$_2$ is much higher
than that in BaFe$_2$As$_2$. A quantitative estimation of the
plasma frequency will be given below. There is only a minor change
in R($\omega$) as temperature decreases from 300 K to 150 K.
However, by cooling across the phase transition at 130 K,
R($\omega$) shows a dramatic change. Except for the reflectance at
very low frequency, R($\omega$) is suppressed significantly below
the reflectance edge, but enhanced in the frequency range of 9000
to 16000 \cm. With further decreasing temperature, the spectral
change becomes very small. The very low frequency R($\omega$)
increases slightly, reflecting enhanced metallic dc conductivity.
In the mean time, a weak suppression in the mid-infrared region
near 5000 \cm could be resolved.

The above mentioned spectral change in R($\omega$) across the
transition is very different from BaFe$_2$As$_2$ which shows an
SDW order concomitant with the a structural transition occurs at
140 K. For BaFe$_2$As$_2$, the most prominent spectral change is a
substantial suppression in R($\omega$) at low frequency, roughly
below 1000 \cm, which is ascribed to the opening of an SDW gap in
the magnetic ordered state\cite{Hu}. In BaNi$_2$As$_2$, the
spectral change occurs over a much broader frequency range.
Besides the week suppression near 5000 \cm, there is no additional
spectral suppression at lower energies.

\begin{figure}[t]
\includegraphics[width=8.5cm,clip]{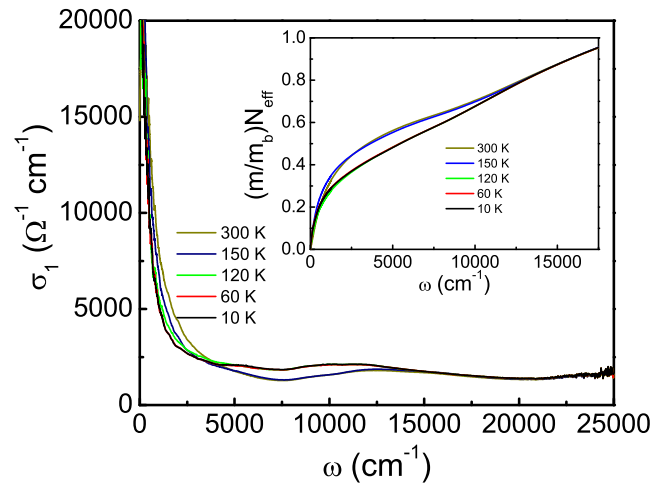}
\caption{\label{fig:R}(Color online) Optical conductivity spectra
over broad frequencies. Inset shows the effective carrier density
per Ni site estimated from the spectral weight under conductivity
curves.}
\end{figure}

Figure 2 shows the conductivity spectra $\sigma_1(\omega$) up to
25000 \cm. The Drude-like conductivity can be observed for all
spectra at low frequencies. When temperature drops below 130 K,
part of the spectral weight transferred from below 7000 \cm to
higher energies. The effective carrier density per Ni site below a
certain energy $\omega$ can be obtained by the partial sum rule
\begin{equation}
  \frac{m}{m_b}N_{eff}(\omega)=\frac{2mV_{cell}}{\pi{e^{2}}N}\int_{0}^{\omega}\sigma_1(\omega')d\omega',
\end{equation}
where m is the free-electron mass, m$_b$ the averaged
high-frequency optical or band mass, $V_{cell}$ a unit cell
volume, N the number of Ni ions per unit volume. The inset of Fig.
2 shows $(m/m_b)N_{eff}$ as a function of frequency for different
temperatures. We find that the spectral weight recovered roughly
near 12000 \cm. $N_{eff}$ is related to an overall plasma
frequency, after choosing a proper high-frequency limit
$\omega_c$, via the relationship
$\omega_p^2=4\pi{e^{2}}N_{eff}(\omega_c)/m_b(V_{cell}/N)=8\int_{0}^{\omega_c}\sigma_1(\omega')d\omega'$.
Choosing $\omega_c\approx$7500 \cm, a frequency where
$\sigma(\omega)$ reaches its minimum but below the interband
transition, we get the overall plasma frequency
$\omega_p\approx$3.49$\times10^4$ \cm for T=300 K and 150 K. If we
choose the same cutoff frequency at 7500 \cm, we get
$\omega_p\approx$3.33$\times10^4$ \cm for T=10, 60 and 120 K.
However,it should be noted that the conductivity spectra already
show a week peak feature near 5000 \cm, this results in an
overestimate of the plasma frequency at low temperatures. Note
that the plasma frequency for BaFe$_2$As$_2$ is only around
1.3$\times10^4$ \cm in the nonmetallic phase\cite{Hu}, so
BaNi$_2$As$_2$ has much higher plasma frequency. This is in accord
with the presence of two more electrons in 3d orbitals.

\begin{figure}[t]
\includegraphics[width=7cm,clip]{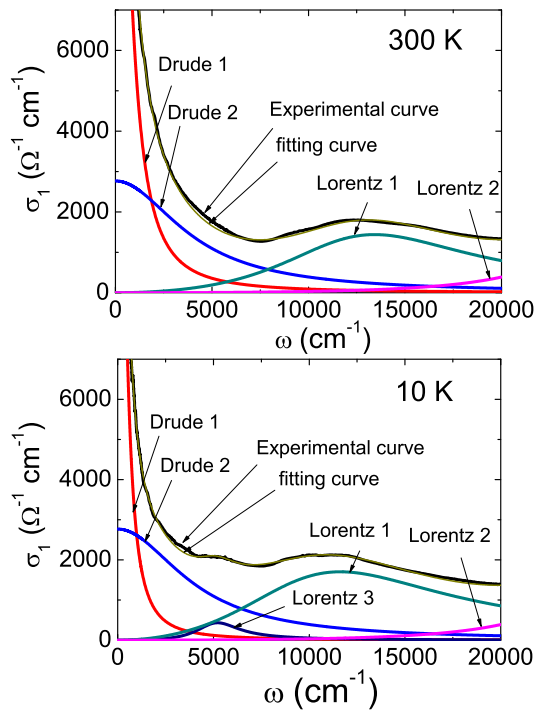}
\caption{\label{fig:R}(Color online) Optical conductivity spectra
at 300 K and 10 K together with the fitting curves from the
Drude-Lorentz model.}
\end{figure}

An alternative way of making quantitative analysis is to fit the
conductivity spectra by a Drude-Lorentz model to isolate the
different components of the electronic excitations\cite{Hu}. The
Drude component represents the contribution from conduction
electrons, while the Lorentz components describe the interband
transitions. Usually a single Drude term is used to extract the
conduction electron contribution, but here we found that two Drude
components could reproduce the low-frequency conductivity much
better. This could be naturally accounted for by the multiple-band
characteristic of the material. The general formula for the
Drude-Lorentz model is
\begin{equation}
\sigma_1(\omega)= \sum{{\omega_{pi}^2 \over4\pi}
{\Gamma_{Di}\over\omega^2 + \Gamma_{Di}^2}} + \sum{{S_j^2
\over4\pi}{\Gamma_j\omega^2 \over (\omega_j^2-\omega^2)^2 -
\omega^2\Gamma_j^2}}, \label{chik}
\end{equation}
where $\omega_{pi}$ and $\Gamma_{Di}$ are the plasma frequency and
the relaxation rate of each conduction band, while $\omega_j$,
$\Gamma_j$, and $S_j$ are the resonance frequency, the damping,
and the mode strength of each Lorentz oscillator, respectively. It
is found that the conductivity curve above the phase transition
could be well reproduced by a two Drude components and two Lorentz
oscillators. Below the transition temperature, an additional
Lorentz oscillator is required to fit the curve. Figure 3 shows
the spectra and fitting curves at two representative temperatures
300 K and 10 K. The parameters for the two Drude components are
summarized in Table 1 for different temperatures. The overall
plasma frequency is obtained as
$\omega_p$=$\sqrt{\omega_{p1}^2+\omega_{p2}^2}$. We find that the
values are roughly in agreement with those obtained by the
sum-rule analysis.

\begin{table}
\caption{\label{tab:table1}The fitting parameters of two Drude
components in Eq. (2) at different temperatures. The unit for
$\omega_{pi}$ and $\Gamma_{Di}$ is \cm.}
\begin{ruledtabular}
\begin{tabular}{cccccc}
 $  T (K)   $&$   \omega_{p1}   $
&$   \Gamma_{D1}   $&$   \omega_{p2}   $&$   \Gamma_{D2}   $&$ \omega_p    $\\
\hline
  $  10    $&$   2.10\times10^4   $&$   410   $&$   2.59\times10^4   $&$   4050   $&$ 3.33\times10^4   $\\
  $  60    $&$   2.10\times10^4   $&$   410   $&$   2.59\times10^4   $&$   4050   $&$ 3.33\times10^4   $\\
  $  120    $&$   2.19\times10^4   $&$   490   $&$   2.59\times10^4   $&$   4050   $&$ 3.39\times10^4   $\\
  $  150    $&$   2.50\times10^4   $&$   510   $&$   2.59\times10^4   $&$   4050   $&$ 3.59\times10^4   $\\
  $  300    $&$   2.74\times10^4   $&$   730   $&$   2.59\times10^4   $&$   4050   $&$ 3.77\times10^4   $\\

\end{tabular}
\end{ruledtabular}
\end{table}

We can now summarize our main findings from the optical data.
BaNi$_2$As$_2$ has a much higher plasma frequency than
BaFe$_2$As$_2$. By lowering the temperature across the transition,
part of the spectral weight from conducting electrons was removed
and shifted to high energies. There is a small peak near 5000 \cm,
but its spectral weight could not compensate for the loss of the
low-frequency Drude component, the spectral weight loss is
recovered at a much higher energy scale. The spectral change
across the transition is very different from the situation of
BaFe$_2$As$_2$, where an SDW gap opens at low frequencies, roughly
below 1000 \cm.\cite{Hu}

The key issue here is the origin of the first order phase
transition in BaNi$_2$As$_2$. As mentioned, a magnetic SDW
transition concomitant with a structural transition was initially
suggested only by analogy with FeAs-based 122 compounds. However,
to date no evidence for magnetic transition was reported.
Furthermore, the band structure calculations reveal that the Fermi
surfaces of BaNi$_2$As$_2$ are markedly different from that of
FeAs-based 122 compounds, the disconnected FS becomes much large
in BaNi$_2$As$_2$\cite{Subedi,Shein}, and no nesting of the Fermi
surfaces could be identified. One would not expect the phase
transitions in these two systems have the same origin. Very
recently, the crystal structure after the transition was
determined. It is found that BaNi$_2$As$_2$ experiences a
structural phase transition from a high-T tetragonal phase to a
low-T triclinic phase. Examining carefully the bondings between
atoms, we find that the As-As distance along the c-axis increases
from 3.5531 $\AA$ to 3.5838 $\AA$, while within ab-plane, the Ni
ions change from a square lattice to zigzag chains where short
Ni-Ni bonds ($\sim 2.8 \AA$) are separated by longer Ni-Ni
distances($\sim 3.1 \AA$)\cite{Sefat}.

\begin{figure}[t]
\includegraphics[width=8.5cm,clip]{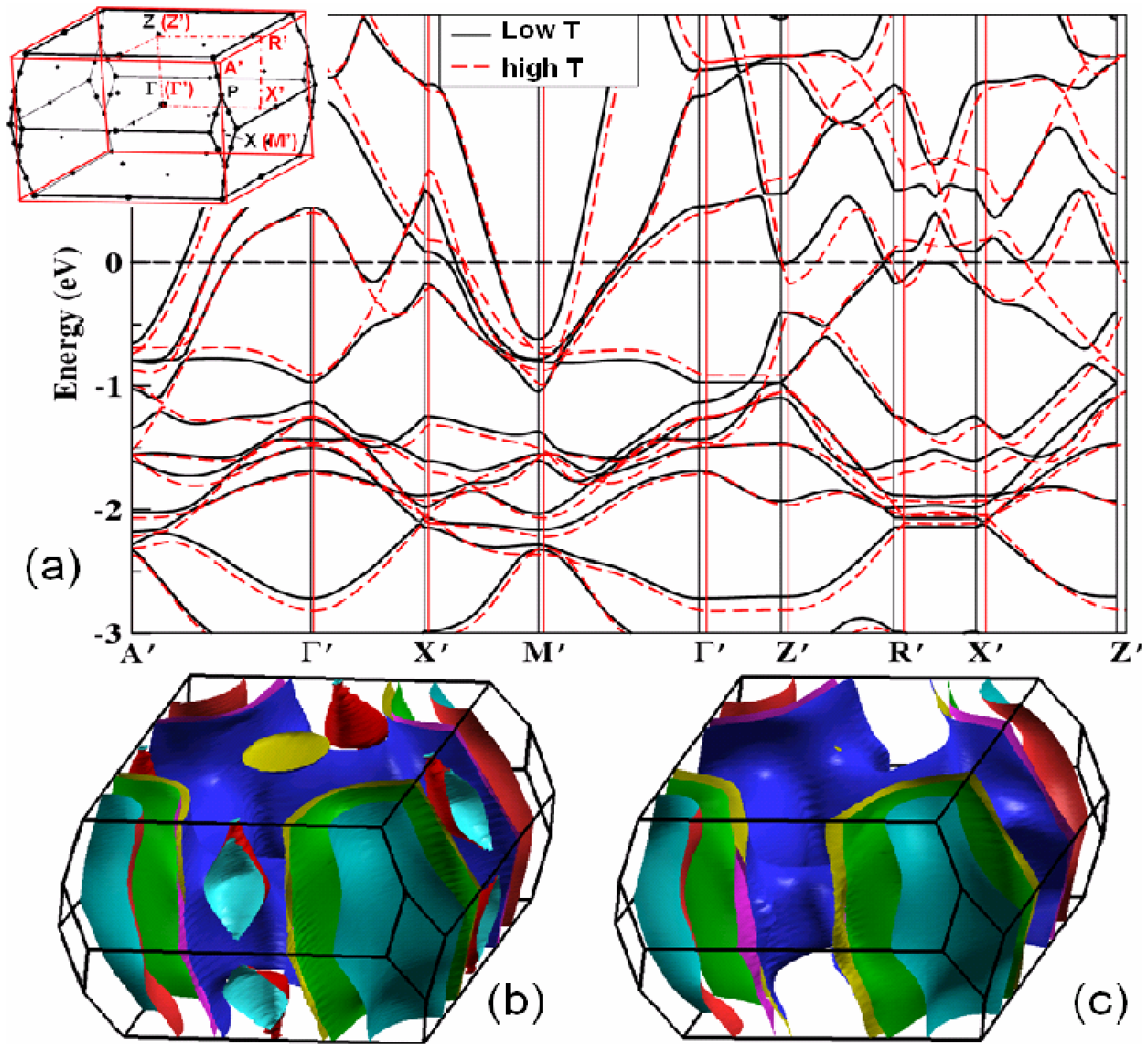}
\caption{\label{fig:R} (a) The band structures for BaNi$_2$As$_2$
both above (red-dash) and below (black) the phase transition. (b)
and (c): The Fermi surfaces in the high-T and low-T phases,
respectively.}
\end{figure}

For the ThCr$_2$Si$_2$-type transition metal-based compounds
AM$_2$X$_2$, three different types of bondings, M-X, M-M, and X-X,
and their contributions to different bands in the electronic
structure were well studied.\cite{Hoffmann,Johrendt} It is found
that two dispersive bands which cross the E$_F$ lightly and have
mainly the X-X bonding (X-X $pp\sigma$) and M-X anitbonding (M-X
$dp\sigma*$) character are rather sensitive to the positions of X
ions, and show strong electron-phonon coupling
effect.\cite{Johrendt} To see the effect of the lattice distortion
on the band structure, we performed LDA band structure
calculations for BaNi$_2$As$_2$ using the experimentally
determined crystal structures below and above the phase transition
as an initial input and then relax the atom positions. We find
that the change of the optimized structure from the experimentally
determined one is rather small, less than 2$\%$ in both atom
positions and lattice parameters. A lowering of energy of 6
meV/unit cell is obtained for the distorted structure. The
calculated results are presented in Fig. 4. Here, a standard
notation of the Brillouin zone for a simple tetragonal lattice is
used for illustrating the band dispersions. For comparison reason,
we also map the Brillouin zone for the low-T triclinic lattice to
the original tetragonal phase. A minor shift in the high symmetry
line is due to the lattice parameter difference. We find that most
of the dispersive bands keep unchanged except for some small
shifts in energy, but the band dispersions close to Fermi level
near Z$^\prime$ point and along R$^\prime$-X$^\prime$ direction
are dramatically altered. Those bands were dominantly contributed
from the As-As $pp\sigma$ bonding and Ni-As $dp\sigma*$
anitbonding. In the high-T symmetry phase, those bands cross the
E$_F$ and lead to a very flat electron type Fermi surface (FS)
surrounding Z$^\prime$ point and eight 3D hole type FS along R-X
line (see Fig. 4 (b)). However, in the low-T phase, the FS near
Z$^\prime$ point tends to vanish almost completely, leading to a
gap roughly about 0.5 eV. Additionally, the hole type 3D FS along
R$^\prime$-X$^\prime$ line is completely gone due to the band
splitting(Fig. 4 (c)). This splitting has smaller energy scale
0.2$\sim$0.3 eV. This could give the onset of the small
mid-infrared peak shown in lower panel of Fig. 3. Compared with
the rather big FS sheets by other bands, the areas enclosed by
those FSs are rather small, so the reduction of the carrier
density below the transition is small, which is also in agreement
with the optical measurement results.

We can see that the driving mechanism for the structural phase
transition is still electronic. The energy gain through gapping
the small FS sheets is not due to the nesting effect of FS, but
mainly caused by the removal/instability of several specific bands
near E$_F$ whose energy levels are particularly sensitive to the
direct As-As and Ni-As bondings. Therefore, our combined optical
spectroscopy and first principle calculation studies firmly
establish that the first-order phase transition in BaNi$_2$As$_2$
is different from that of BaFe$_2$As$_2$.

\begin{acknowledgments}
This work is supported by the National Science Foundation of
China, the Chinese Academy of Sciences, and the 973 project of the
Ministry of Science and Technology of China.
\end{acknowledgments}


\end{document}